\documentclass[cits,hyper]{PoS}
\newcommand{\mres}{m_\textrm{res}}

\title{The finite temperature phase transition from domain wall fermions}

\ShortTitle{The finite temperature phase transition from domain wall fermions}

\author{\speaker{Michael Cheng}\thanks{Current Address: Center for Computational Science, Boston University, Boston, MA 02215} for the HotQCD Collaboration\\
        Lawrence Livermore National Laboratory\\
		7000 East Avenue\\
		Livermore, CA 94550\\
        E-mail: \email{cheng24@llnl.gov}}

\abstract{We present results on the finite temperature QCD transition with 2+1 flavors using Domain Wall Fermions (DWF) with the Dislocation Suppressing Determinant Ratio (DSDR).  In particular, we discuss how the use of DSDR allows us to study the finite temperature transition at the coarse lattice spacings corresponding to the transition region ($T = 139 - 195$ MeV) with DWF at $N_\tau = 8$.  The residual chiral symmetry breaking at these lattice spacings is sufficiently small so that a constant pion mass of $m_\pi \approx 200$ MeV is obtained in our calculations.  The strange quark mass is set to near its physical value.  We show results on the restoration of chiral symmetry and deconfinement at finite temperature.
}

\FullConference{ The XXIX International Symposium on Lattice Field Theory - Lattice 2011\\
July 10-16, 2011\\
Squaw Valley, Lake Tahoe, California}

\begin{document}

\section{Introduction}

The finite temperature transition for QCD at zero baryon density has been a subject of intense study for the
lattice gauge theory community.  The nature of the phase transition and the value
for the pseudo-critical temperature $T_c$ is not only of fundamental interest, but also has phenomenological
implications for experimental studies of relativistic heavy ion collisions, as well as the evolution of the
early universe.

Most large-scale studies of the QCD transition on the lattice have used staggered fermions.
Staggered fermions are computationally inexpensive, and preserve a $U(1)$ remnant of the
$SU(2)_L \otimes SU(2)_R$ chiral symmetry at finite lattice spacing.  However, lattice artifacts in the
staggered formulation break flavor symmetry, leading to, \textit{e.g.} artificial mass splittings in
the hadron spectrum.

It would be desirable to use a lattice action that more faithfully reproduces the continuum
symmetries of QCD.  One such action is known as domain wall fermions (DWF).  DWF possesses an exact
$SU(2) \otimes SU(2)$ chiral symmetry, even at finite lattice spacing.  This is accomplished by adding
a fictitious fifth dimension into the lattice action, where the physical chiral modes are bound
to opposite four-dimensional walls.  If the fifth dimension is infinitely large,
the chiral modes are isolated and the chiral symmetry is exact.  However, for any real lattice calculation,
one must choose a finite fifth-dimensional size, $L_s$.  In such a calculation, $L_s$ 
controls the amount of mixing between the chiral modes, and thus the residual
chiral symmetry breaking.

There have been some past studies of chiral symmetry breaking with domain wall fermions \cite{Chen:2000zu,Cheng:2009be}.  However,
in these past studies, it has been difficult to control the residual chiral symmetry breaking, 
which increases very quickly as one moves towards strong coupling.  In this work, we present a new
study of the transition region with domain wall fermions, employing a modified gauge action, the
Dislocation Suppressing Determinant Ratio (DSDR) with the Iwasaki action on the gauge links.
This combination  keeps residual chiral symmetry breaking small, allowing us to study the
transition region while keeping the physical pion mass fixed, $m_\pi \approx 200$ MeV.

\section{Method}

\subsection{Dislocation Suppressing Determinant Ratio (DSDR)}

In our lattice calculation, we use domain wall fermions, along with the DSDR method and the Iwasaki gauge
action.  The finite fifth dimension results in a residual chiral symmetry breaking caused by a mixing
of the chiral modes between the four-dimensional walls.  To leading order, this results in a simple
additive renormalization to the quark masses, the residual mass $\mres$.  The residual mass
may be parameterized by \cite{Antonio:2008zz}:
\begin{equation}
\mres = c_1 \rho_H(\lambda_c)\frac{e^{-\lambda_c L_s}}{L_s} + c_2 \rho_H(0)\frac{1}{L_s},
\label{eq:mres}
\end{equation}
where $\rho_H(\lambda)$ denotes the eigenvalue density of the effective four-dimensional
Hamiltonian.  The first term in eqn. \ref{eq:mres} comes from de-localized states with eigenvalues near
the mobility edge, $\lambda_c$, while the second term comes from localized "lattice dislocations" with near
zero eigenvalue.  As one moves to strong coupling, the rapid proliferation of these localized
dislocations means that the latter term in eqn. \ref{eq:mres} dominates the residual mass for
moderate values of $L_s \sim 16-32$.  In this regime, $\mres \sim 1/L_s$, meaning that it becomes
very inefficient to reduce the chiral symmetry breaking by increasing the size of the fifth dimension.

The near-zero eigenmodes of the effective Hamiltonian are closely related to
the near-zero eigenmodes of the the Hermitian Wilson-Dirac operator, $H_5 = \gamma^5 D_W(-M_0)$, where
$M_0$ is the domain wall height.  Thus, it has been suggested that augmenting the action with
the determinant of $H_5$ will suppress exactly those modes that make the biggest contribution
to $\mres$ \cite{Vranas:2006zk,Fukaya:2006vs}.  However, it is exactly these near-zero modes of $H_5$
that allow topology to change during a molecular dynamics evolution.  In order to allow for topology
change during our Monte Carlo calculation, we instead introduce the following weight factor into our
action:
\begin{eqnarray}
\label{eqn:DSDR}
\mathcal{W}(M_0, \epsilon_b, \epsilon_f) & = & \frac{\det\left[D^\dagger_W(-M_0 + i \epsilon_f \gamma^5)D_W(-M_0 + i \epsilon_f \gamma^5)\right]}{\det\left[D^\dagger_W(-M_0 + i \epsilon_b \gamma^5)D_W(-M_0 + i \epsilon_b \gamma^5)\right]}\\
& = & \frac{\det\left[D^\dagger_W(-M_0)D_W(-M_0) + \epsilon^2_f\right]}{\det\left[D^\dagger_W(-M_0)D_W(-M_0) + \epsilon^2_b\right]}\nonumber,
\end{eqnarray}
where $\epsilon_b$ and $\epsilon_f$ are chirally-twisted mass terms.  By tuning $\epsilon_b$ and $\epsilon_f$
appropriately, we can find a weighting factor that reduces $\mres$ sufficiently without completely eliminating
topology change during our HMC evolution.  The choice of $\epsilon_f = 0.02$ and $\epsilon_b = 0.50$ satisfies these conditions \cite{Renfrew:2009wu}. We call the weighting factor, $\mathcal{W}(M_0, \epsilon_b, \epsilon_f)$
the Dislocation Suppressing Determinant Ratio (DSDR).  

\subsection{Line of constant physics}

Using the DSDR method, we can choose our input bare light quark masses $m_l$ such that the total 
quark mass $m_\textrm{tot} = m_l + \mres$ corresponds to a fixed physical pion mass, $m_\pi \approx 200$
MeV.  We have produced configurations at 7 different temperatures with spatial size $N_\sigma = 16$ and
temporal extent $N_\tau = 8$, spanning the temperature range $T \in [139, 195]$ MeV.  
This corresponds to lattice spacings of $a \approx 0.13 - 0.18$ fm.  For each ensemble, we
have produced between 2996-7000 molecular dynamics trajectories.
Table \ref{tab:finite} summarizes the parameters that we use in our finite temperature ensembles.

\begin{figure}[hbt]
\begin{minipage}[hbt]{0.45\textwidth}
\begin{center}
\includegraphics[width=\textwidth]{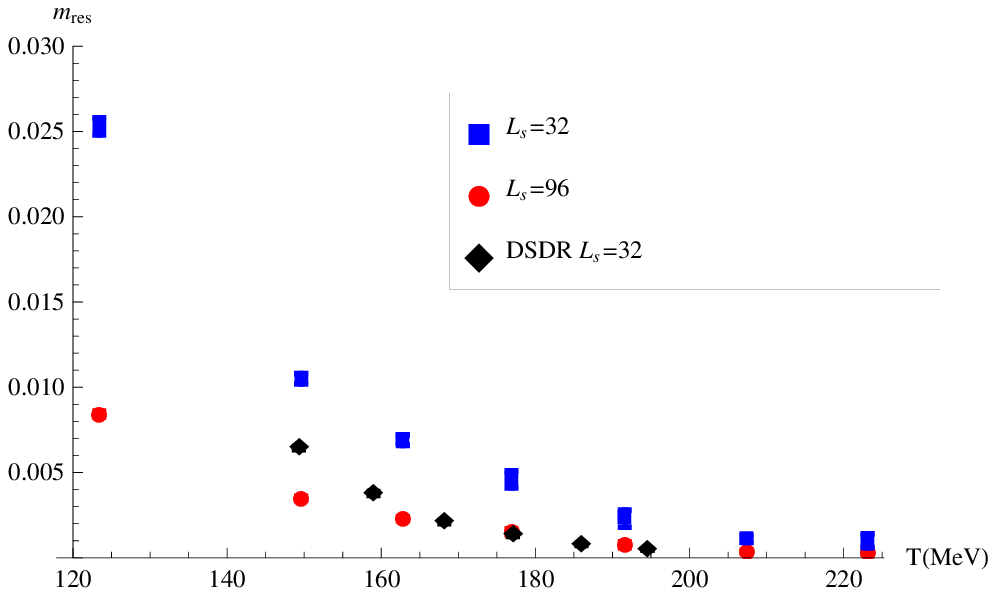}
\caption{Comparison of $\mres$ from domain wall fermions without the DSDR method at $L_s = 32, 96$ and
what is obtained using DSDR at $L_s = 32$.}
\label{fig:mres_DSDR}
\end{center}
\end{minipage}
\hspace{0.07\textwidth}
\begin{minipage}[hbt]{0.45\textwidth}
\begin{center}
\includegraphics[width=\textwidth]{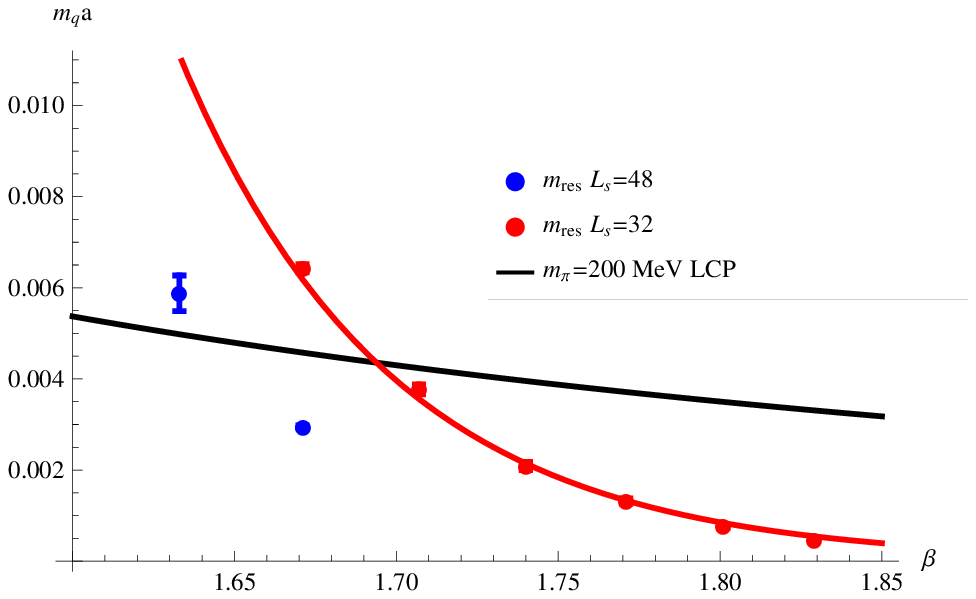}
\caption{The solid black curve plots the total bare quark mass $m_\textrm{tot}$ required for a fixed physical pion mass $m_\pi = 200$ MeV as a function of the bare coupling $\beta$.  The red curve and points show $\mres$ for the $L_s = 32$ DSDR lattices.  The blue points show $\mres$ for the $L_s = 48$ ensembles.}
\label{fig:mres_comp}
\end{center}
\end{minipage}
\end{figure}

Fig. \ref{fig:mres_DSDR} shows a comparison of $\mres$ as a function of temperature between
our current calculation with domain wall fermions and the DSDR method, compared to domain wall
fermions without the DSDR method, both at $N_\tau = 8$.  With $L_s = 32$, the DSDR method
offers a substantial improvement in residual chiral symmetry breaking.  In fact, for temperatures
$T > 170~\textrm{MeV}$, one needs to have at least $L_s = 96$ in order to achieve residual masses equal
to what can be obtained using DSDR at $L_s = 32$.

In order to determine the lattice scales in our calculation and the quark masses needed in our
finite temperature ensembles, we have also used results from zero temperature calculations at
three values of the gauge coupling $\beta  = 1.70, 1.75, 1.82$ that lie within our range of
interest.  Table \ref{tab:zero} summarizes the parameters for our zero temperature ensembles.

Using interpolations and extrapolations from the zero temperature data, we are able to determine
the total bare light quark masses, $m_\textrm{tot} = m_l + \mres$ required to keep the physical
pion mass fixed to $m_\pi = 200~\textrm{MeV}$.  Fig. \ref{fig:mres_comp} shows this value of 
$m_\textrm{tot}$ as a function of the bare coupling, $\beta$.  Also shown are the values of $\mres$
obtained on the DSDR finite temperature lattices with $L_s = 32$ and $L_s = 48$.  Even with $L_s = 48$,
the lowest temperature ensemble ($\beta = 1.633$) requires a negative input light quark mass.  At 
$\beta = 1.671$ we have generated two separate ensembles with $L_s =32$ and $L_s = 48$, with
a negative and positive input light quark mass, respectively, in order to test
if the negative quark mass produces any deleterious effects.  With these parameters, we found
 no evidence of any
"exceptional configurations" that are, in principal, possible with a negative quark mass.

\begin{table}[hbt]
\begin{center}
\begin{tabular}{cccccccccc}
\multicolumn{10}{c}{Finite Temperature Ensembles}\\
$T$ (MeV)   &$\beta$&$L_s$&$m_l a$&$m_s a$& $\mres$ &$m_\pi$ (MeV) &$\chi_{l,disc}/T^2$ & $\chi^{\overline{MS}}_{l.disc}/T^2$ &Traj.\\
\hline
139(6)&1.633&48&-0.00136&0.0519& 0.00588(39) &191(7)&37(3) & 17.2(14) &2996 \\
 149(5)&1.671&32&-0.00189&0.0464& 0.00643(9) & 199(5)&44(3) & 19.9(10) & 6000\\
 149(5)&1.671&48&0.00173 &0.0500& 0.00295(3) &202(5) & 41(2) & 18.5(9) &7000 \\
 159(4)&1.707&32&0.000551&0.0449& 0.00377(11) &202(3) & 43(4) & 18.8(18) &3659 \\
 168(4)&1.740&32&0.00175 &0.0427& 0.00209(9) &197(2) & 35(5) & 14.9(21) & 3343 \\
 177(4)&1.771&32&0.00232 &0.0403& 0.00132(6) &198(2) & 25(4) & 10.4(17) & 3540 \\
 186(5)&1.801&32&0.00258 &0.0379& 0.00076(3) &195(3) & 11(4) & 4.5(1.6) &4715 \\
 195(6)&1.829&32&0.00265 &0.0357& 0.00047(1) &194(4) & 5(3) & 2.0(1.2) & 6991 \\
\hline
\end{tabular}
\end{center}
\caption{Summary of finite temperature ensembles with DSDR and results for the disconnected chiral susceptibility, $\chi_{l,disc}$.}
\label{tab:finite}
\end{table}

\begin{table}[hbt]
\begin{center}
\begin{tabular}{ccccccccccc}
\multicolumn{11}{c}{Zero Temperature Ensembles}\\
\hline
$\beta$&$N_\sigma$&$N_\tau$&$L_s$&$m_l a$&$m_s a$& $\mres$ &$m_\pi$ (MeV) & $r_0/a$ & $a^{-1}$ (GeV) &Traj.\\
\hline
1.70 & 16 & 32 & 32 & 0.013 & 0.047 & 0.00420(2) &394(9) & 2.895(11) & 1.27(4) & 1360 \\
1.70 & 16 & 32 & 32 & 0.006 & 0.047 & 0.00408(6) &303(7)& 2.992(27) & 1.27(4) & 1200 \\
1.75$^*$ & 32 & 64 & 32 & 0.0042 & 0.045 & 0.00180(5) &246(5)& 3.349(20) & 1.36(3) & 1288 \\
1.75$^*$ & 32 & 64 & 32 & 0.001 & 0.045 & 0.00180(5) &172(4)& 3.356(22) & 1.36(3) & 1560 \\
1.82 & 16 & 32 & 32 & 0.013 & 0.040 & 0.00062(2) &398(9) &3.743(28) & 1.55(5) & 2235 \\
1.82 & 16 & 32 & 32 & 0.007 & 0.040 & 0.00063(2) &304(7)&	 3.779(37) & 1.55(5)& 2134 \\
\hline
\end{tabular}
\end{center}
\caption{Summary of zero temperature ensembles with DSDR.  Lattice scales determined using $r_0 = 0.487(9)$ fm, after extrapolation to the chiral limit.
$^*$: $\beta = 1.75$ results are RBC-UKQCD preliminary results. }
\label{tab:zero}
\end{table}

\section{Chiral Symmetry Restoration}

On our finite temperature lattice ensembles we have computed the connected, spatial two-point functions
in various mesonic channels, $G_\Gamma(x)$.  From these, we can trivially calculate the corresponding
connected susceptibilities,
\begin{equation}
\frac{\chi^{con}_\Gamma}{T^2} = N_\tau^2 \sum_x G_\Gamma(x).
\label{eqn:chi_con}
\end{equation}

We have also measured the one-flavor scalar chiral condensate, $\left<\overline{\psi}_l \psi_l\right>$
and the pseudoscalar condensate, $\left<\overline{\psi}_l \gamma^5 \psi_l\right>$ with a stochastic estimator.
Using these, we can compute the disconnected scalar and pseudoscalar susceptibilities,
\begin{eqnarray}
\frac{\chi^{disc}_l}{T^2} & = & N_\sigma^3 N_\tau^3 \left(\left<(\overline{\psi}_l \psi_l)^2\right>-\left<\overline{\psi}_l \psi_l\right>^2\right)\\
\frac{\chi^{disc}_5}{T^2} & = & N_\sigma^3 N_\tau^3 \left(\left<(\overline{\psi}_l \gamma^5 \psi_l)^2\right>-\left<\overline{\psi}_l \gamma^5 \psi_l\right>^2\right)
\label{eqn:chi_disc}
\end{eqnarray}

With the connected and disconnected susceptibilities, we can trivially reconstruct the flavor singlet ($\sigma$) and non-singlet ($\delta$) susceptibilities in the scalar channel,
\begin{equation}
\frac{\chi_\sigma}{T^2}  = \frac{\chi^{con}_l + \chi^{disc}_l}{T^2};~~~
\frac{\chi_\delta}{T^2}   =  \frac{\chi^{con}_l}{T^2},
\label{eqn:chi_scalar}
\end{equation}
and also the flavor singlet ($\eta$) and non-singlet ($\pi$) susceptibilities in the pseudoscalar channel,
\begin{equation}
\frac{\chi_\eta}{T^2} =  \frac{\chi^{con}_5 - \chi^{disc}_5}{T^2};~~~
\frac{\chi_\pi}{T^2}  =  \frac{\chi^{con}_5}{T^2}.
\label{eqn:chi_pseudo}
\end{equation}

\begin{figure}[t]
\begin{minipage}[hbt]{0.45\textwidth}
\hspace{-0.10\textwidth}
\includegraphics[width=\textwidth,angle=-90]{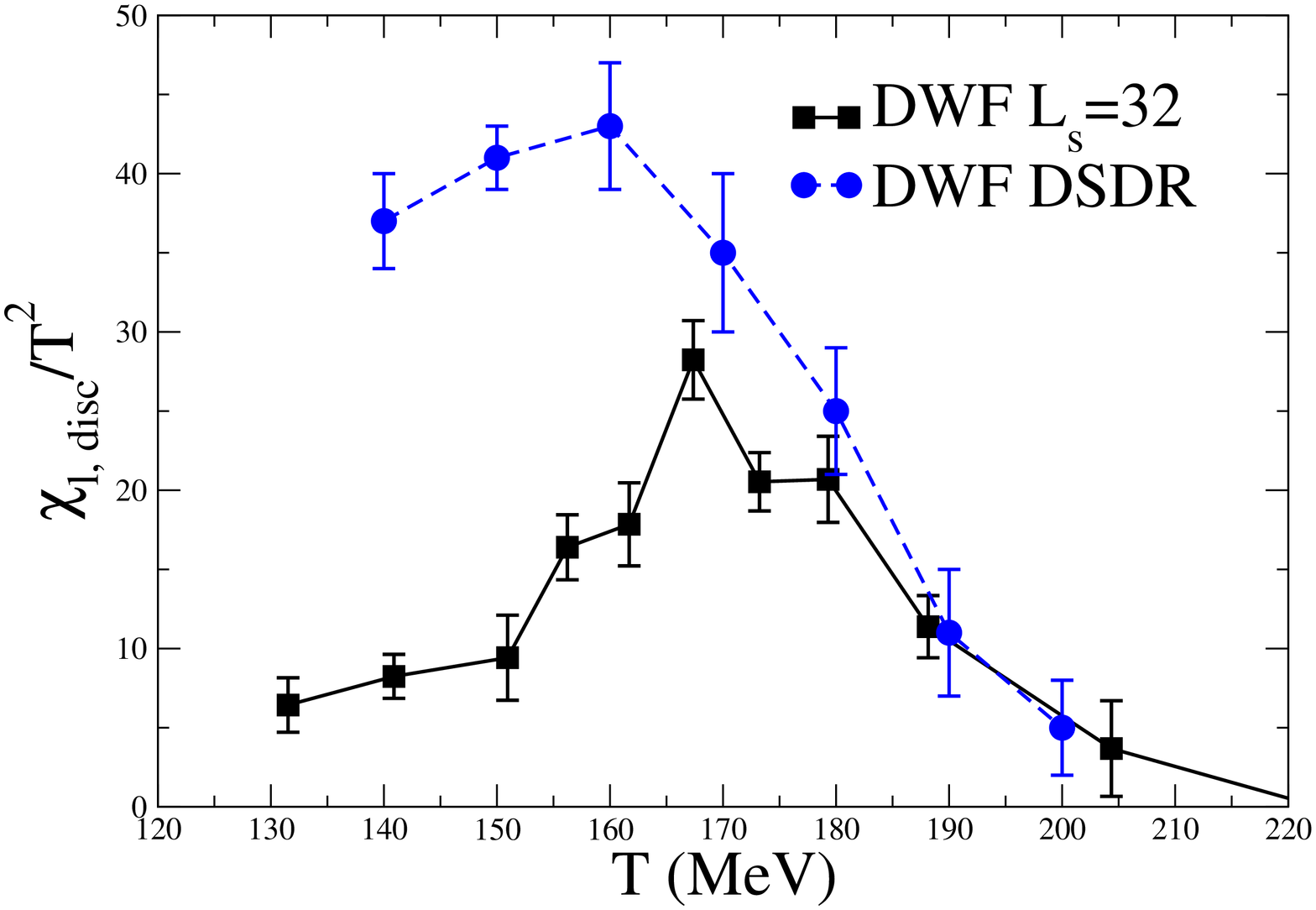}
\caption{Comparison of the disconnected chiral susceptibility in our calculation (DWF DSDR) with earlier
results \cite{Cheng:2009be} (DWF $L_s=32$), where the finite temperature ensembles do not lie along a line of constant physics.}
\label{fig:DWF_comp}
\end{minipage}
\hspace{0.07\textwidth}
\begin{minipage}[hbt]{0.45\textwidth}
\hspace{-0.10\textwidth}
\includegraphics[width=\textwidth,angle=-90]{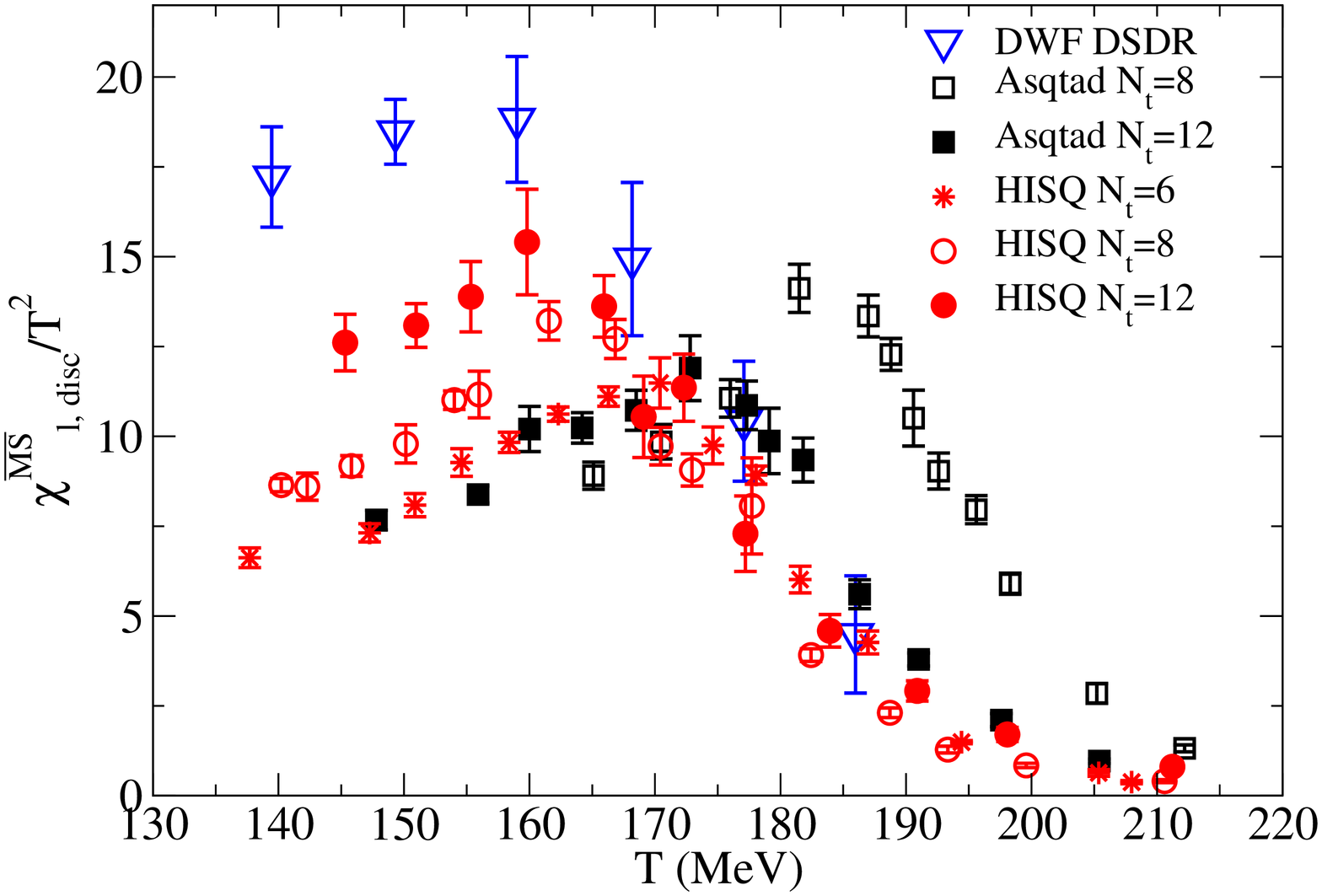}
\caption{Comparison of the disconnected chiral susceptibility with results from various staggered actions \cite{Bazavov:2011nk}.  All
results have been converted to the $\overline{\textrm{MS}}$ renormalization scheme.}
\label{fig:staggered_comp}
\end{minipage}
\end{figure}

One of the signatures for chiral symmetry restoration is the divergence of the disconnected chiral
susceptibility ($\chi^{disc}_l$) at $T_c$, along with the corresponding $O(4)$ scaling near $T_c$, in the chiral
limit.  Thus, one method to identify the crossover temperature at finite quark mass is by the peak
in the chiral susceptibility.  Fig. \ref{fig:DWF_comp} is a comparison of the disconnected chiral susceptibility
in our current calculation with the earlier results from \cite{Cheng:2009be}, which were not performed along
a line of constant physics.  Although there is a clear peak near $T \approx 170$ MeV in the earlier results, 
the results in the low temperature region are significantly distorted compared to our current calculation. 
In our calculation, a peak is visible near $T \approx 160$ MeV.  

Fig. \ref{fig:staggered_comp} compares our results with the results of various improved staggered actions \cite{Bazavov:2011nk}.
In order to compare results from different actions, all results are converted into the $\overline{\textrm{MS}}$
renormalization scheme.  While the lightest pions in the staggered calculation correspond to a physical quark mass 
that is approximately half as light as in our calculation, there seems to be reasonably good agreement for 
$\chi^{disc}_l$ for $T > T_c$.
However, whereas the location of the peak in $\chi^{disc}_l$, $T_c \approx 160$ MeV is approximately
the same as for the HISQ action with $N_\tau = 12$, the magnitude of the susceptibility for $T \lesssim T_c$
is significantly greater for DWF than the various staggered actions.  This may be due to 
finite volume effects, as the DWF calculation has an an aspect raio $N_\sigma/N_\tau = 2$, compared to 
$N_\sigma/N_\tau = 4$ for the staggered calculations.

In the chirally-restored phase, we can also derive relations between
the scalar and pseudoscalar susceptibilities.  The chiral transformation mix the scalar and pseudoscalar
channels, so that in the high temperature phase, we have the relations $\chi_\pi = \chi_\sigma$ and
also $\chi_\delta = \chi_\eta$.  Thus, we expect that the differences in these susceptibilities should
vanish in the chiral limit at temperatures where chiral symmetry is restored.
\begin{figure}[h]
\begin{minipage}[hbt]{0.45\textwidth}
Using the definitions in eqns.
\ref{eqn:chi_scalar} and \ref{eqn:chi_pseudo}, we can also show that the
difference in the pseudoscalar and scalar disconnected susceptibilities, 
$\Delta_{disc} = \chi^{disc}_5 - \chi^{disc}_l$ should also vanish in the chirally restored phase.  Fig.
\ref{fig:chi_diff} shows $\Delta_{disc}/T^2$ as well as the difference $(\chi_\pi - \chi_\sigma)/T^2$.
Even though we are not in the chiral limit, both of these differences are already consistent with zero
for $T \approx 170-180$ MeV, suggesting that chiral symmetry is already well restored at those temperatures.
\end{minipage}
\hspace{0.07\textwidth}
\begin{minipage}[hbt]{0.45\textwidth}
\begin{center}
\includegraphics[width=\textwidth]{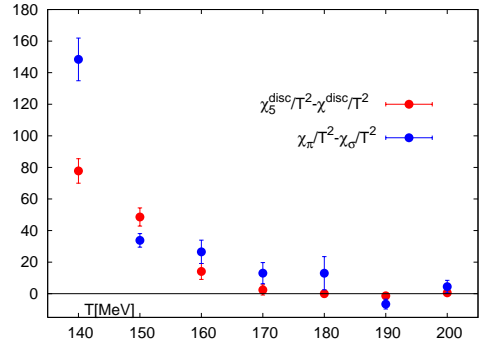}
\caption{Plot of $\Delta_{disc}/T^2 = (\chi^{disc}_5-\chi^{disc}_l)/T^2$ and $(\chi_\pi - \chi_\sigma)/T^2$ as a function of temperature.}
\label{fig:chi_diff}
\end{center}
\end{minipage}
\end{figure}

\section{Conclusion}

In this work, we have examined the restoration of chiral symmetry in finite temperature QCD
using domain wall fermions on lattices of spatial size $N_\sigma = 16$ and temporal extent
$N_\tau = 8$ in the temperature range $T \in [139,195]$ MeV.  In order to control the
residual chiral symmetry breaking,
we have employed the DSDR method, augmenting the gauge action with a weighting factor that suppresses
the localized dislocations that contribute most to $\mres$.  
By controlling residual chiral symmetry breaking, we were able to use a fixed, physical pion mass
$m_\pi \approx 200$ MeV.

In order to find the chirally restored phase, we examined the light disconnected chiral susceptibility,
which exhibits a peak at the crossover temperature $T_c$.  A comparison with an earlier domain wall
calculation, which did not have a fixed pion mass, show the distortions introduced if one does not
do calculations along a line of constant physics.  A comparison with recent results with improved
staggered fermions shows significant differences in the chiral susceptibility, especially for 
$T \leq T_c$.  Another signature of chiral symmetry restoration that we examined were the
scalar and pseudoscalar susceptibilities, both the flavor singlet and non-singlet channels.  
By examining the differences in these susceptibilities,
we could see apparent chiral restoration for $T \approx 170-180$ MeV.

In addition to the restoration of chiral symmetry, we have also examined the effective restoration of
$U(1)_A$ symmetry as well as the eigenvalue spectrum on these finite temperature lattices.  These
results are presented in these proceedings in \cite{Hegde:2011zg,Lin:2011bj}, respectively.
Future calculations are underway to extend these results to larger spatial volume in order to
control finite-volume effects, as well as to smaller quark masses, so that we can examine the
quark mass dependence of these quantities.


\end{document}